%% file: main.tex
\documentclass[a4paper,11pt]{article}
\pdfoutput=1 

\usepackage{jinstpub} 

\usepackage{lineno}

\usepackage{booktabs, multirow} 
\usepackage{soul}
\usepackage[table]{xcolor} 
\usepackage{changepage,threeparttable} 
\usepackage{siunitx}

\title{
Development of a novel bunch oscillation recorder with RFSoC technology
}

\author[1]{R. Nomaru,}
\author[2]{G. Mitsuka,}
\author[3]{L. Ruckman,}
\author[3]{and R. Herbst}
\affiliation[1]{
    The University of Tokyo, Bunkyo, Tokyo 113-0033, Japan
}
\affiliation[2]{
    KEK, Oho, Tsukuba, Ibaraki 305-0801, Japan
}
\affiliation[3]{
    SLAC National Accelerator Laboratory,
    2575 Sand Hill Road, M/S 96,
    Menlo Park, CA 94025, USA
}

\emailAdd{nomaru@g.ecc.u-tokyo.ac.jp}

\abstract{

\noindent

The SuperKEKB accelerator is designed to achieve unprecedented luminosity levels, but this goal is currently hindered by Sudden Beam Loss (SBL) events. These events not only obstruct luminosity improvement but also pose a significant risk to accelerator components, the Belle II detectors, and the superconducting focusing system, potentially leading to severe damage and quenching of the superconducting system. To address this critical challenge, we have developed a novel Bunch Oscillation Recorder (BOR) based on RFSoC technology. The BOR has demonstrated high precision with a position resolution of 0.03 mm, making it a powerful tool for real-time beam monitoring. In its initial deployment, the BOR successfully recorded multiple SBL events, providing valuable data for further analysis. By strategically positioning BORs at the suspected points of SBL origin, we aim to directly identify sources of beam instability. We anticipate that this portable, high-speed BOR monitor will play a crucial role in resolving the SBL issue, ultimately helping achieve SuperKEKB's luminosity targets.
}

\keywords{
Beam diagnostics; 
RFSoC; 
Particle accelerator;
Instrumentation
}

\begin{document}
\maketitle
\flushbottom
\setlength{\parskip}{0pt}


\input{00_Background}
\input{01_Hardware}

\input{02_Firmware}
\input{03_Software}
\input{04_Testing}

\input{05_Analysis}
\input{06_summary}

\input{bibliography}
\end{document}

%% file: 00_Background.tex
\section{Background and Motivation}
\label{sec:Background}

The SuperKEKB accelerator~\cite{SuperKEKBTDR} collides electrons and positrons at very high luminosity, providing large quantities of $B$ mesons, $\tau$ leptons, and other charged particles for the Belle II detector. The accelerator consists of a $\SI{7}{\giga\electronvolt}$ electron storage ring (HER) and a $\SI{4}{\giga\electronvolt}$ positron storage ring (LER). By adopting a nanobeam scheme, the beam size at the collision point was narrowed down to a nanometer level, and the world's highest luminosity record, $\SI{4.7e+34}{\per\square\centi\meter\per\second}$, was achieved in 2022. The accelerator resumed operations in February 2024 after its first long shutdown of over a year. We are operating towards even higher luminosity, but a phenomenon called Sudden Beam Loss (SBL) is hindering luminosity improvement~\cite{Ikeda:2023evb}.

\begin{figure}[hbt]
 \centering
 \includegraphics[width=0.7\columnwidth]{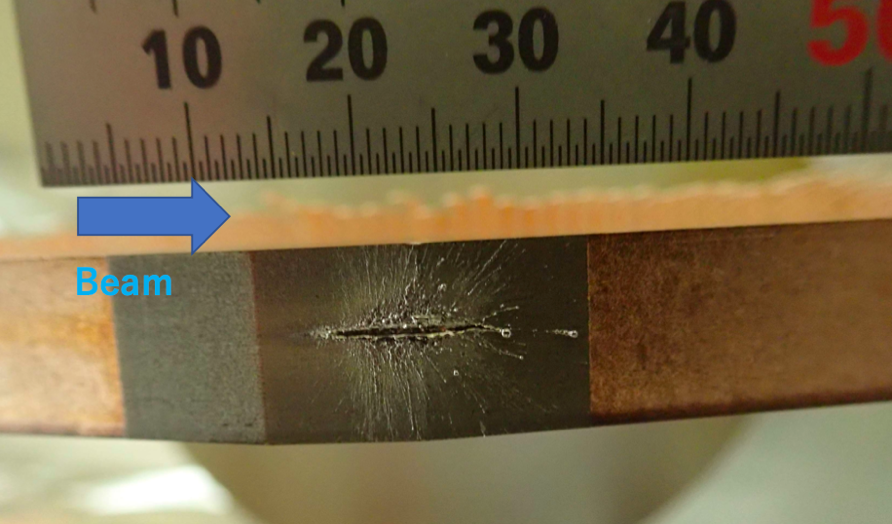}
 \caption{Damaged collimator head due to the sudden beam loss event in LER.}
 \label{fig:collimator}
\end{figure}

The SBL event is presently the biggest obstacle to increasing the integrated luminosity at SuperKEKB. 
SBL events have caused cracks in accelerator components such as collimators, damage to the Belle II detector, and quenches in the superconducting focusing system. 
Figure~\ref{fig:collimator} shows how the SBL events have damaged a tantalum-made collimator head~\cite{Ishibashi}. 
Note that 31 collimators in both rings protect the Belle II detector from unwanted beam losses.
Collimator head replacement involves removing damaged and scattered head material (tungsten or tantalum) from the vacuum pipe and restoring the pressure in the vacuum pipe to the ultra-high vacuum level after it has deteriorated to the atmosphere level during the replacement process. 
Therefore, the entire replacement process usually takes several days to a week.
We find, mostly empirically, that the higher the stored bunch intensity, the more frequently the SBL events occur, often followed by severe damage to accelerator components.
Thus, we hesitate to increase the stored beam current as anticipated, hindering the increase in instantaneous luminosity. 
In addition, once the SBL events occur, leading to beam aborts or the superconducting system quenching, recovery to regular beam operation takes several hours, up to an entire day. 
Therefore, since SBL events limit both, increasing the instantaneous luminosity and gaining integrated luminosity as well, the events have become a critical issue in SuperKEKB operations. 
Nevertheless, the cause of the SBL event and the location of its occurrence have yet to be identified.

\begin{figure}[hbt]
 \centering
 \includegraphics[width=0.85\columnwidth]{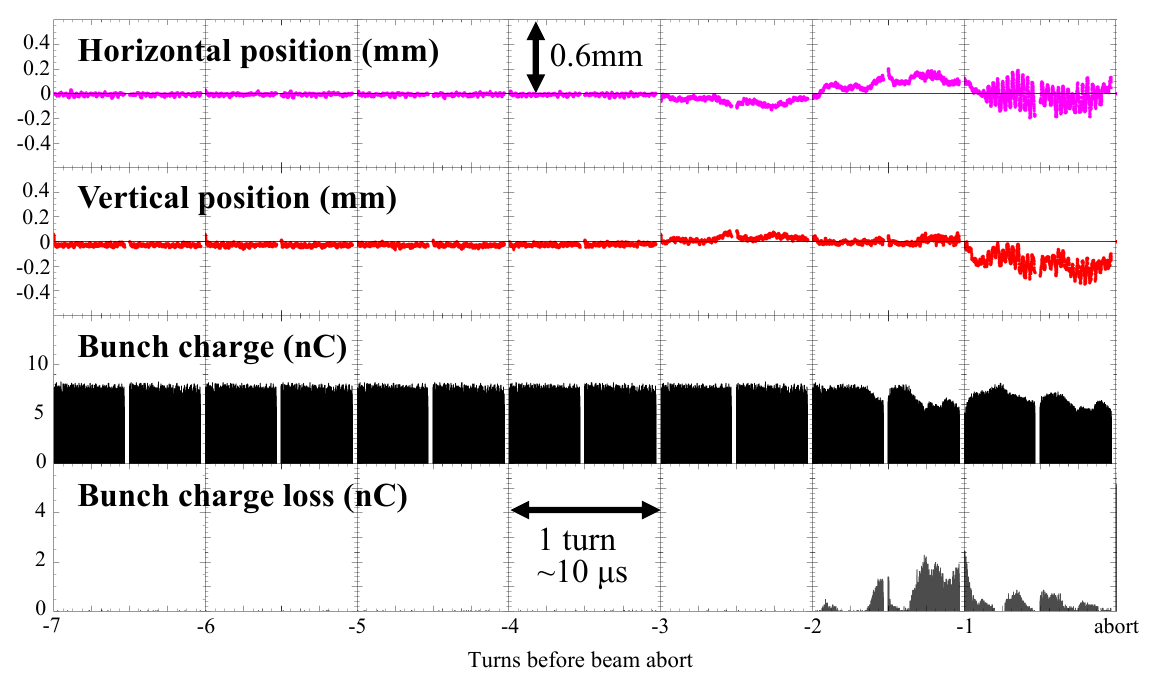}
 \caption{Example of the SBL event. This SBL occurred at the LER on March 23, 2022. Bunch-by-bunch horizontal orbit, vertical orbit, bunch current, and bunch current loss from the top. Each turn has two bunch trains, each with 636 bunches.}
 \label{fig:SBL_VMEBOR}
\end{figure}

A Bunch Oscillation Recorder (BOR) is a beam instrument that records the beam position bunch-by-bunch for ten or more turns before a beam abort occurs. 
Before the first long shutdown, a Versa Module Eurocard (VME) type BOR, 18K10 by Digitex Co., Ltd, hereafter we call it VME-BOR, was used for each ring~\cite{Tobiyama:IBIC12-MOPA36}. 
The position measurement accuracy of this VME-BOR is approximately \SI{30}{\micro\meter}.
VME-BORs in each ring have been placed in the Fuji straight section of the SuperKEKB main ring and acquired bunch-by-bunch oscillation data before every beam aborts. 
Measurements with the VME-BOR indicate that the SBL events are accompanied by sizable bunch position oscillations.
Figure~\ref{fig:SBL_VMEBOR}~\cite{Michele} shows how the horizontal and vertical bunch positions move for three turns before the beam abort triggered by a beam loss of $\sim\SI{10}{\percent}$.
Each graticule division in Figure~\ref{fig:SBL_VMEBOR} corresponds to one turn.
Such observation indicates an unknown source, potentially residual dust, beam instabilities, and an imperfect feedback kicker system that kicks a portion of bunches, leading to bunch position oscillation and significant beam loss. 
However, only one VME-BOR in each ring cannot pinpoint where and how SBL events initially occur. 
It cannot distinguish two possible scenarios: small kick amplitudes or degeneracy of the betatron phase advance. 
Thus, we realize the necessity of placing multiple BORs to cover phase advances widely and more directly detect the bunch oscillation near suspicious incidence locations. 
The VME-BOR design is over ten years old, and the many parts used inside the circuit, such as the 8-bit ADC, are unavailable today. 
In light of this situation, our aim in the project is to develop a "handy" BOR with position measurement accuracy comparable to or even higher than that of the VME-BOR ($\sim\SI{30}{\micro\meter}$), ready for operation after the first long shutdown and place several copies of it along the ring.
In recent years, advancements in digital circuit technology have made the development of bunch-by-bunch monitors a hot topic in the accelerator research field \cite{BbB_paper,ipac_nsls}.
As a benchmark for future bunch-by-bunch monitors, we developed this BOR using a new chip called RF System on Chip made by AMD/Xilinx.

%% file: 01_Hardware.tex
\section{Hardware}
\label{sec:hardware}

\subsection{Overview}
\begin{figure}[hbt]
 \centering
 \includegraphics[width=\columnwidth]{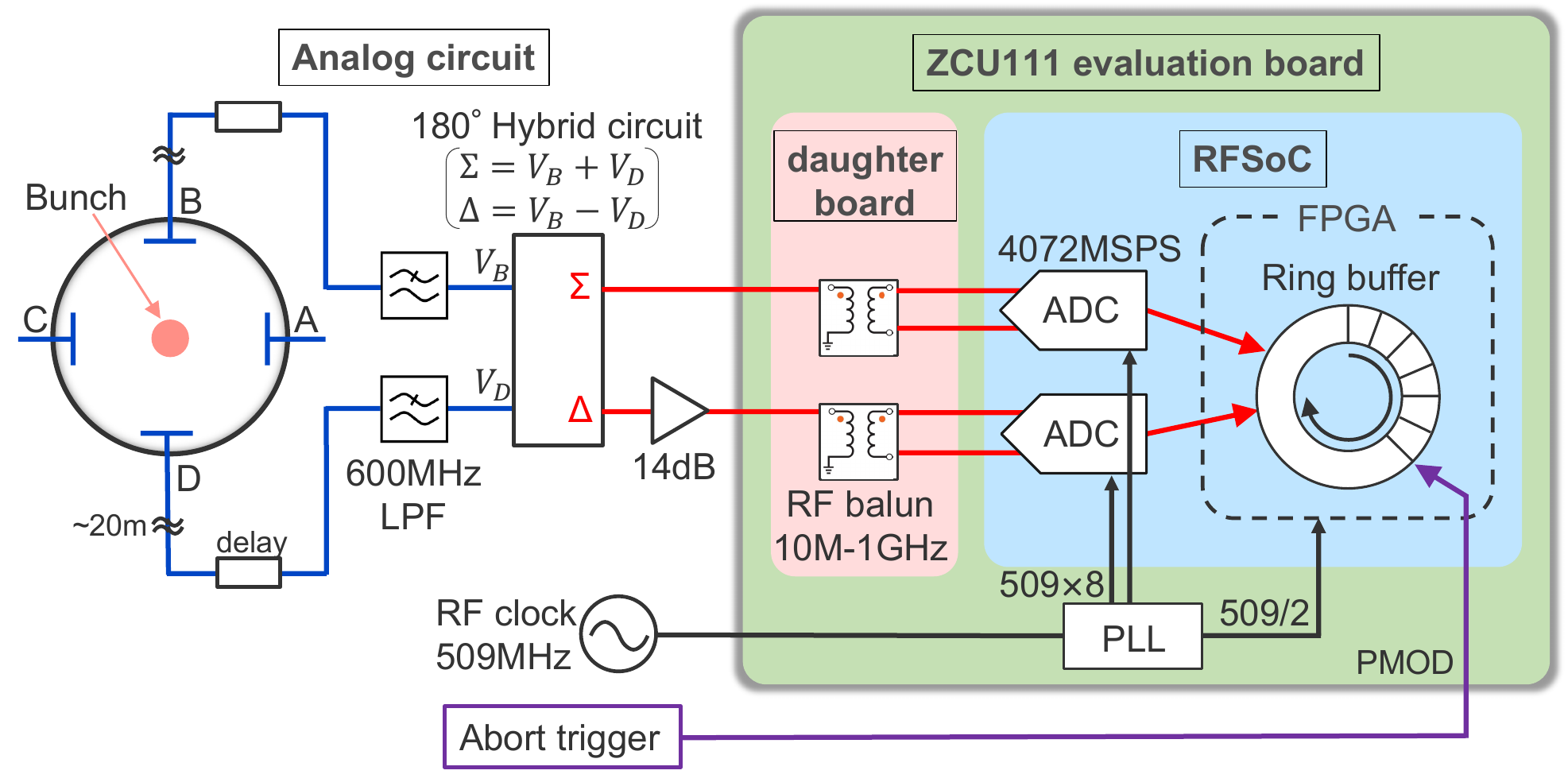}
 \caption{Schematic diagram of the RFSoC-based BOR system}
 \label{fig:schematic_of_BOR}
\end{figure}

The new BOR was developed using a novel integrated chip solution referred to as an RF System on Chip (RFSoC), which was developed by AMD/Xilinx. 
The RFSoC has been acknowledged as a significant advancement in the fields of radio frequency and digital signal processing. 
Data converters, programmable logic, and processor are integrated on a single monolithic chip, referred to as RFSoC, designed to streamline the development of high-performance radio frequency systems \cite{WP489,rfsoc}. 
This R\&D effort was undertaken using the ZCU111 RFSoC evaluation board \cite{zcu111}. 
Signals from the button electrodes of the beam position monitor (BPM) pickup in the accelerator vacuum chamber were input with some minimum signal conditioning and analog pre-processing to the RFSoC, sampled and acquired to measure the position of bunches. 
Figure~\ref{fig:schematic_of_BOR} shows the schematic diagram of the RFSoC-based BOR system we developed.

\subsection{Vacuum chamber and button electrode}
\begin{figure}[hbt]
 \centering
 \includegraphics[width=\columnwidth]{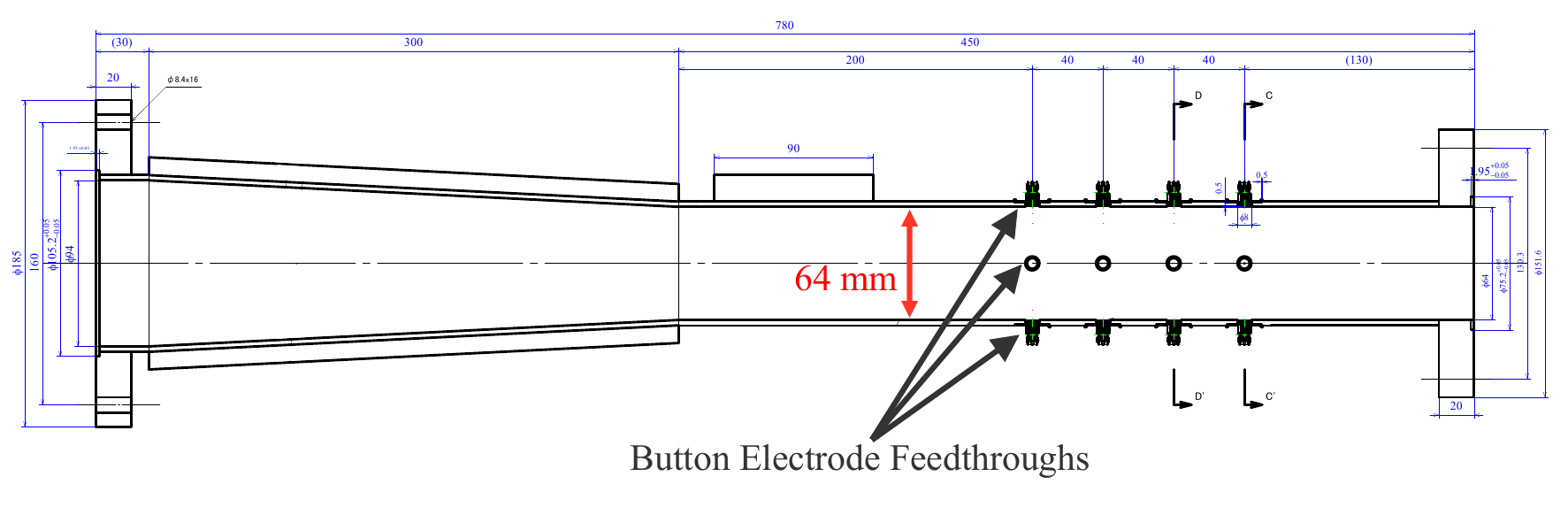}
 \caption{Design of the SuperKEKB FB-BPM chamber}
 \label{fig:chamber}
\end{figure}

\begin{figure}[hbt]
 \centering
 \includegraphics[width=\columnwidth]{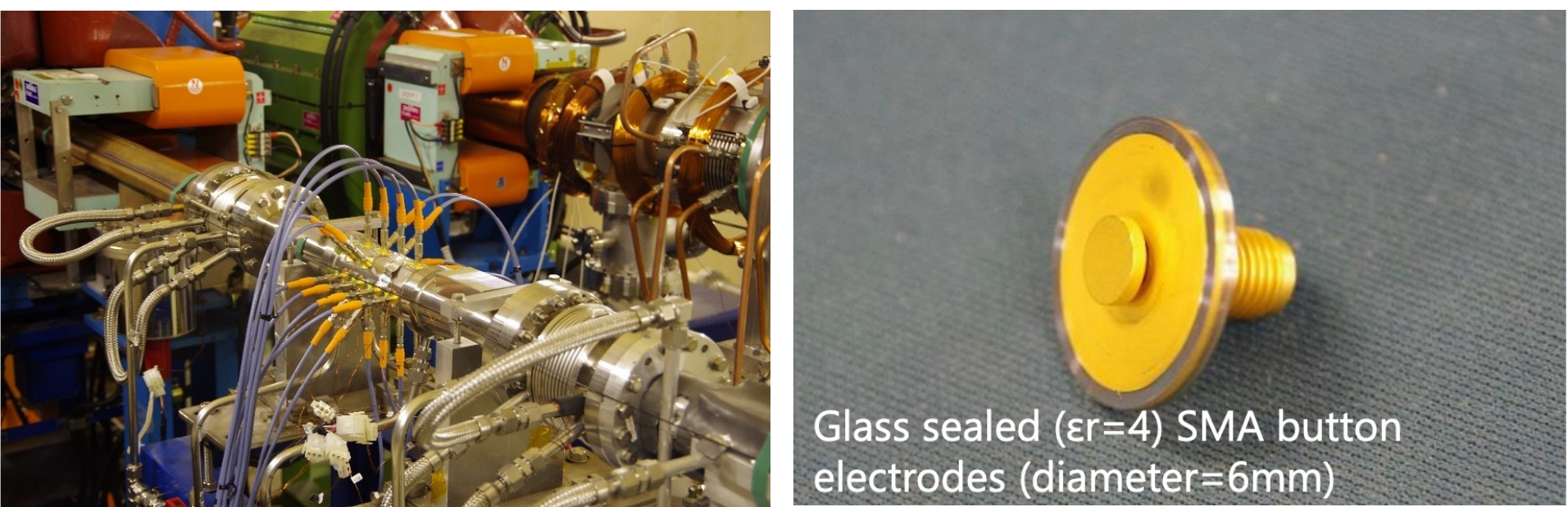}
 \caption{(Left) The SuperKEKB FB-BPM chamber. The cable visible in the center of the photo is connected to the button electrode inside the beam pipe. (Right) The button electrode attached to the FB-BPM chamber. This has a diameter of \SI{6}{\milli\meter}, is sealed with glass, and has an SMA connector.}
 \label{fig:button_electrode}
\end{figure}

Figure~\ref{fig:chamber} shows the design of the vacuum chamber used for this R\&D. 
This is a bunch-by-bunch orbit feedback BPM (FB-BPM) chamber \cite{chamber}, installed in the Fuji straight section of the SuperKEKB accelerator. 
The diameter of the chamber in the area where the button BPM is installed is \SI{64}{\milli\meter}.
Figure~\ref{fig:button_electrode} (Left) shows a picture of this FB-BPM chamber, which is equipped with a total of 24 button electrodes.
Four button electrodes are connected to the bunch-by-bunch orbit feedback. The other electrodes are available for detector R\&D. In this experiment, we used the two vertically oriented electrodes to measure the vertical position.
Figure~\ref{fig:button_electrode} (Right) shows a photo of a BPM button electrode, as it is mounted in the FB-BPM chamber. 
The button electrode has a diameter of \SI{6}{\milli\meter}, is sealed with glass, and features an SMA connector \cite{chamber,electrode}.

We can obtain the vertical bunch position $y$ using the ratio of the two voltage signals of the BPMs facing each other ($V_B$ and $V_D$ in Figure~\ref{fig:schematic_of_BOR}). The vertical bunch position is formulated as
\begin{equation}
\label{position_eq}
y=k_y\frac{V_B-V_D}{V_B+V_D}=k_y\frac{\Delta}{\Sigma}
\end{equation}
where $k_y$ is a constant calculated by the boundary element method \cite{boundary} and $\Delta=V_B-V_D, \Sigma=V_B+V_D$. 
The FB-BPM chamber has the button electrodes arranged in a symmetric, horizontal/vertical cross pattern, therefore the bunch position can be determined using the voltages of only two electrodes facing each other.

\subsection{Analog circuit}
\label{analog_circuit}
\begin{figure}[hbt]
 \centering
 \includegraphics[width=0.7\columnwidth]{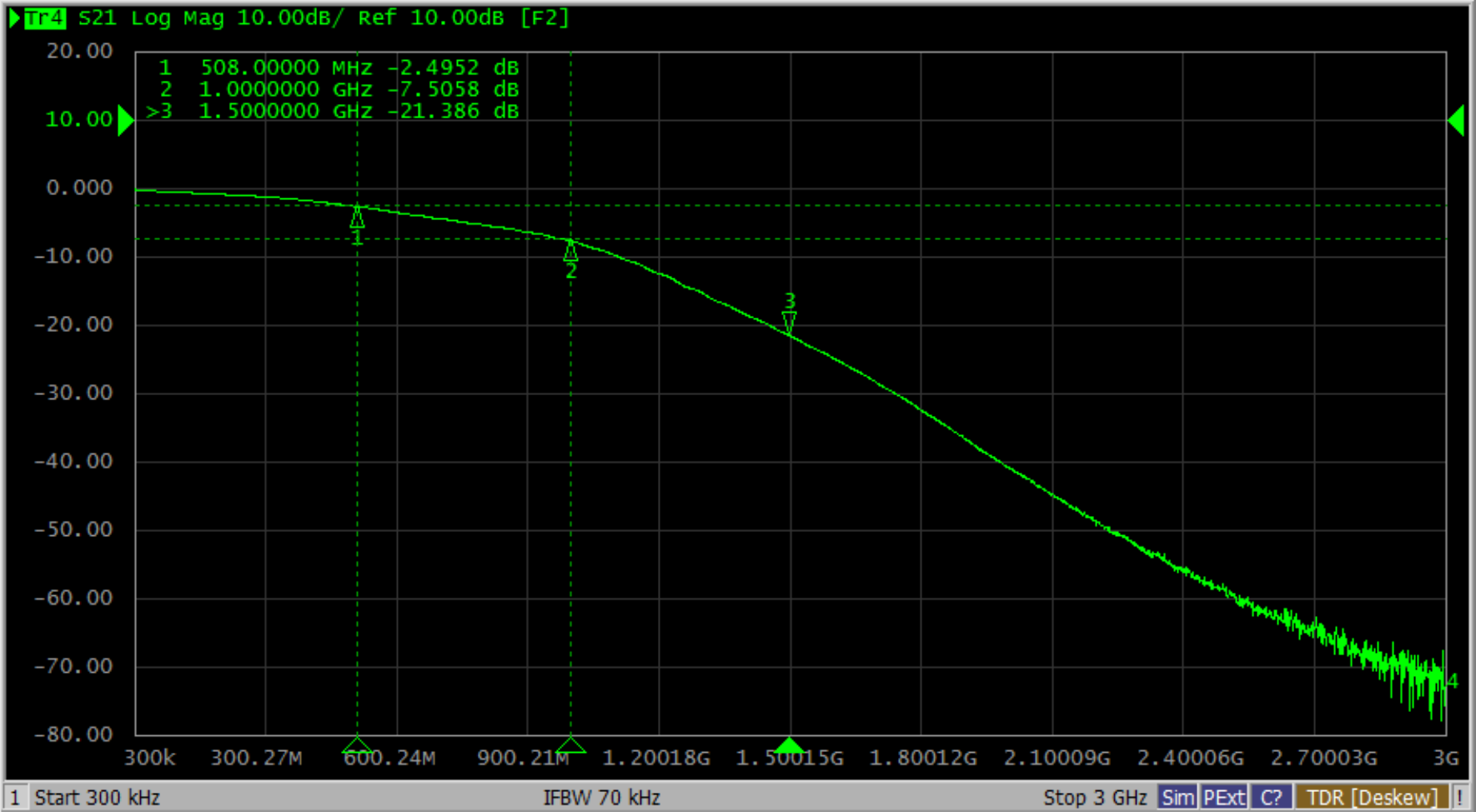}
 \caption{Frequency response of low-pass filter 5LP8-600B-SR by LORCH. This is the $|S_{21}(f)|$ transfer parameter of this low-pass filter measured with a network analyzer.}
 \label{fig:LORCH_response}
\end{figure}

The signals of the vertical $B$ and $D$ button electrodes are transmitted by coaxial cables of approximately 20~m in length to a radiation-free counting room outside the accelerator tunnel, where all the signal processing hardware is located.
To compensate for different signal arrival times due to cable length mismatch, two coaxial airlines of variable length, so-called "trombone" phase shifters made by Nihon Koshuha Co., Ltd \cite{nihon_koshuha}, were used to align the timing of signals from the two button electrodes.
Following this, the signals passed through a 600 MHz low-pass filter (LPF) to remove high-frequency ringing from the beam signal.
We selected the commercial LPF, type 5LP8-600B-SR made by LORCH \cite{LORCH}, from a variety of filters we tested, as it provided the best performance in terms of minimum ringing and tails of the output waveform, while still separating the bunch signals, spaced by \SI{4}{\nano\second}.
The frequency response of this filter is shown in Figure~\ref{fig:LORCH_response}. 
This is the $|S_{21}(f)|$ transfer parameter of this LPF measured with a vector network analyzer (VNA), type E5071C, made by Keysight Technologies \cite{network_analyzer}.
It validates the proper function with a 3dB cutoff frequency of 600 MHz, as specified by the vendor.

\begin{figure}[hbt]
 \centering
 \includegraphics[width=\columnwidth]{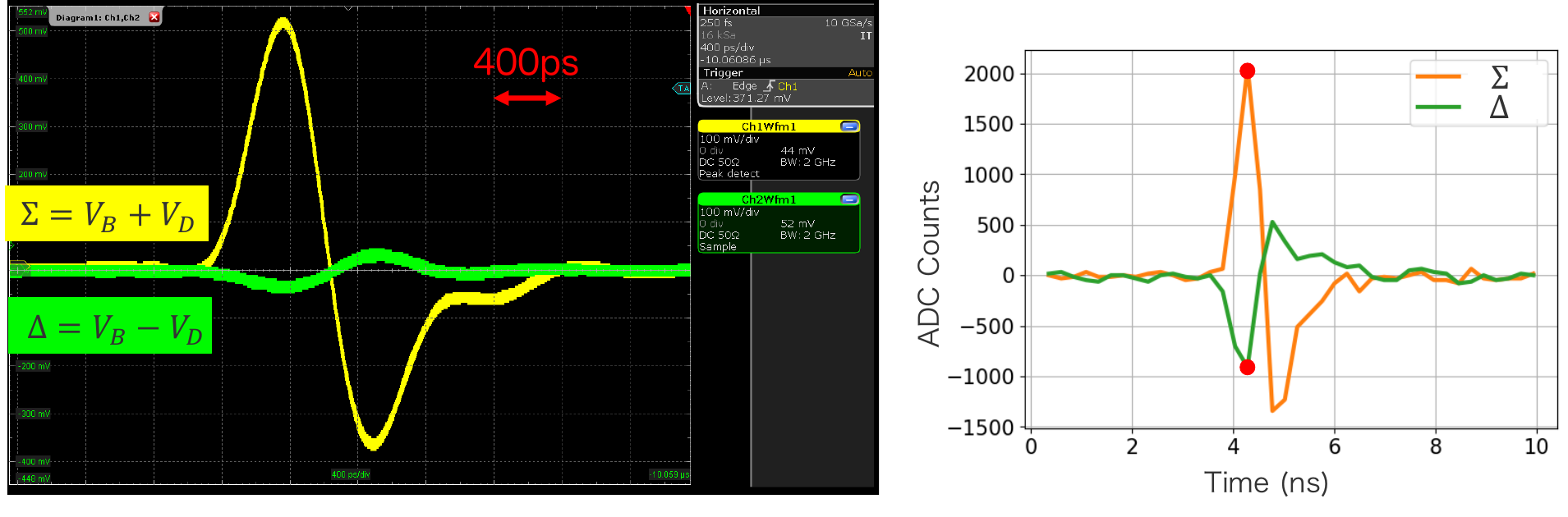}
 \caption{(Left) Hybrid circuit output measured by an oscilloscope. The yellow waveform is $\Sigma$ and the green waveform is $\Delta$. This shows the signal from one bunch in the LER with an intensity of 10~nC. (Right) The digital signal recorded by the RFSoC with the sampling frequency of 4.072~GSPS. The orange waveform is $\Sigma$ and the green waveform is $\Delta$. This shows the signal from one bunch in the LER with an intensity of 2~nC.}
 \label{fig:waveform}
\end{figure}

The outputs of the two LPFs were then fed into a 180-degree hybrid circuit to extract the sum signal "$\Sigma$" and the difference signal "$\Delta$" in eq. (\ref{position_eq}). 
We used the hybrid circuit H-183-4-SMA, made by MACOM \cite{H-183-4}. In the 100-1500~MHz range, H-183-4-SMA has the insertion loss of 1.5~dB maximum, the isolation is 20~dB minimum, and the VSWR is 1.6:1 maximum. 
Figure~\ref{fig:waveform} (left) shows the output signals of the hybrid circuit measured by an oscilloscope. 
The yellow waveform is $\Sigma$, and the green waveform is $\Delta$. 
The oscilloscope is RTO2024, made by Rohde \& Schwarz, with a sampling speed of 10~GSPS and a bandwidth of 2~GHz \cite{oscilloscope}. 
The signal in Figure~\ref{fig:waveform} (left) represents the hybrid output signals for a bunch intensity of about 10~nC. The waveform data was recorded on February 8, 2024, where the bunch spacing was \SI{6}{\nano\second} or more. 

As the $\Delta$ signal was small, it was passed through an amplifier after the hybrid circuit. 
We used the THS4303EVM amplifier evaluation board made by Texas Instruments \cite{THS4303EVM}. 
The THS4303 amplifier chip \cite{THS4303} implemented on this evaluation board has a fixed gain and is a single-ended input/output amplifier.
It has a gain of 20~dB and a bandwidth of 1.8~GHz. 
In this evaluation board, a gain of 14 dB was achieved after impedance matching to \SI{50}{\ohm}. 
At a typical bunch intensity of 10~nC, the $\Delta$ signal level was approximately \SI{50}{\milli\volt}, plus a gain of 14~dB resulted to approximately \SI{250}{\milli\volt}, which was well within the input range of the RFSoC's ADC, \SI{1}{\volt} peak-to-peak.
Since the outputs of the hybrid circuit and the inputs of the ZCU111 evaluation board are both single-ended, we used the THS4303EVM amplifier which had the single-ended input and output ports.
$\Sigma$ and $\Delta$ signals were then fed into the ZCU111 evaluation board.

\subsection{ZCU111 evaluation board}
The ZCU111 evaluation board utilizes the first-generation RFSoC, supporting eight channels of 12-bit ADCs (up to 4.096~GSPS) and eight channels of 14-bit DACs (up to 6.554~GSPS). 
Synchronization of the ADCs' sampling frequency and the FPGA clock frequency inside RFSoC with the SuperKEKB RF clock 509~MHz \cite{SuperKEKBTDR} is achieved through an external clock input and phase locked loops (PLLs). 
The ZCU111's ADC inputs and DAC outputs are connected to SMA connectors via an AMD daughter card called XM500 RFMC (Radio Frequency Mezzanine Card) \cite{xm500}. 
Four of the eight ADC inputs are designated for differential signals, while the other four accept single-ended signals. 
Two of the single-ended channels are connected to a TCM2-33WX+ balun made by Mini-Circuits \cite{LF_balun} for 10~MHz to 1~GHz ("Low Frequency"), and the remaining two channels are connected to a BD1631J50100AHF balun made by Anaren \cite{HF_balun} for 1 to 4~GHz ("High Frequency").

For our application, we used the "Low Frequency" (LF) balun inputs of the XM500 daughter board. 
Through the LF balun, the signal was input into the RFSoC and was digitized by the RF-ADC. 
The digitized signals were transmitted to the internal FPGA for post-processing.
The FPGA is equipped with ring buffers that temporarily hold these digital signals. 
Figure~\ref{fig:waveform} (Right) shows the signals from a pair of button electrodes of the FB-BPM chamber in the LER, pre-processed by the analog circuit as described in Section \ref{analog_circuit}, and acquired by the RFSoC with the sampling frequency of 4.072~GSPS. 
The orange waveform is $\Sigma$ and the green waveform is $\Delta$. 
These waveforms were from a single bunch with an intensity of 2~nC, recorded on February 16th, 2024. 
The bunch spacing for this measurement was $\geqq$ \SI{8}{\nano\second}. 
The sampling frequency of 4.072~GSPS was performed by 8-fold multiplication of the SuperKEKB RF frequency of 509~MHz. 
The sampling frequency and phase were correctly synchronized with the SuperKEKB RF frequency to lock one of the sampling points exactly on the peak amplitude of the bipolar waveform signal of successive bunches.
By acquiring data points at the constant time intervals that were multiples of the RF period, it is possible to meats the beam position of all passing bunches at a repetition, i.e. bunch spacing of $\geqq$ \SI{4}{\nano\second}.

The abort kicker trigger signal of SuperKEKB is received via the ZCU111's PMOD (Peripheral Module) GPIO (General Purpose Input/Output) interface. 
This abort kicker trigger is issued when the loss monitors installed throughout the accelerator observe radiation exceeding a certain threshold value and this activates the abort kicker, causing the beam to be aborted. 
When this GPIO receives this signal, it issues a fault trigger to the ring buffer. 
In the current firmware design, the ring buffer stores 100 turns of digitized $\Sigma$ and $\Delta$ signals before the abort time.
This enables a bunch-by-bunch measurement of the beam position behavior for approximately 1~ms before the beam abort.
This is because one turn takes about \SI{10}{\micro\second}, therefore 100 turns correspond to about \SI{1}{\milli\second}.
Finally, we can calculate the bunch position by taking the peak of the bipolar signal for each recorded $\Sigma$ and $\Delta$ (red points in Figure~\ref{fig:waveform} (Right)) and computing their ratio and multiplying by $k_y$ during offline analysis, following eq. (\ref{position_eq}).

%% file: 02_Firmware.tex
\section{Firmware}
\label{sec:firmware}

\begin{figure}[tb]
\centering
\includegraphics[width=1.0\textwidth]{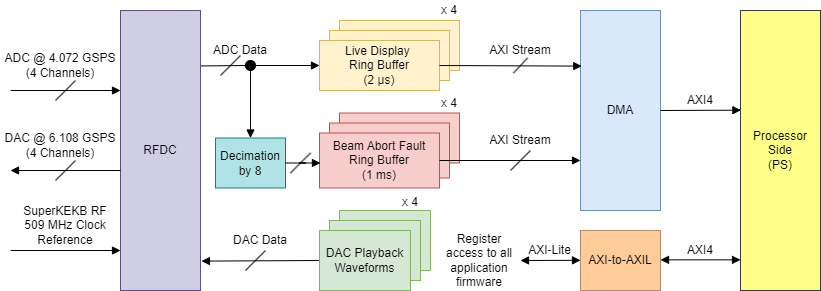}
\caption{\label{fig:fw_block_diagram} Block diagram of the firmware}
\end{figure}

The firmware running on the RFSoC is illustrated in a block diagram in Figure \ref{fig:fw_block_diagram}. 
At the core of the digitization process, the Radio Frequency Data Converter (RFDC) block is positioned as a "hardened IP" within the RFSoC, responsible for the conversion of ADC and DAC data. 
The RFDC block is integrated as a dedicated, fixed component within the RFSoC hardware. In contrast to programmable logic, where IP cores are implemented using FPGA resources and referred to as "soft IP" due to their flexible and reconfigurable nature, hardened IP is pre-designed and optimized for specific functionalities. This integration provides superior performance and reliability, albeit with reduced configurability.
Configured to operate at 4.072~GSPS across 4 channels, the ADCs, capable to clock at a maximum of 4.096~GSPS, receive four analog signals. 
Similarly, the DACs, with a maximum clock rate of 6.554~GSPS, are set to operate at 6.108~GSPS, converting the digital data back into analog signals across 4 channels. 
Sourced from the SuperKEKB RF reference, a 509 MHz reference clock is utilized. 
The internal Phase Lock Loops (PLLs) of the RFDC convert this 509 MHz reference clock into a 4.072 GHz sampling clock for the ADC channels and a 6.108 GHz sampling clock for the DAC channels. 
At a higher integer multiple of the 509 MHz reference clock, the DAC channels are operated faster than the ADC channels to provide improved waveform generation.

The digital data from the ADC channels are stored in two types of ring buffers: live display and fault. 
The live display buffers, which have a depth of two microseconds, are triggered periodically and asynchronously by software to update the waveform display. 
This aids in monitoring and debugging. 
In contrast, the fault buffers, which have a depth of approximately 1 ms, are triggered by a beam abort signal connected to the ZCU111's PMOD interface. 
These fault buffers record the signal at a decimated rate of 8, a necessary reduction to maximize the buffered time span. 
Additionally, the firmware provides the capability to manually trigger the fault buffer from software for monitoring and debugging purposes unrelated to faults. 
The data from the read-out ring buffers is transmitted as an AXI Stream to the Direct Memory Access (DMA) block using the AXI Stream protocol \cite{axis}. 
Subsequently, the DMA transfers the data over an AXI4 interface to the Processor Side (PS) using the AXI4 memory protocol \cite{axil}.

The DAC channels are used to emulate the BPM signals for early firmware and software development when BPM beam signals are unavailable for testing. 
The software can load waveform patterns into the Programmable Logic's (PL's) memory for each DAC channel, making them appear sufficiently similar to the BPM signals of the analog front-end. 
Next, the firmware plays back these loaded waveforms through the RFDC's digital DAC interface. 
This playback can be either continuous or occur for a programmable number of iterations. 
To use this BPM emulation, the DAC SMA connectors and ADC SMA connectors on the XM500 daughter card are connected together via coaxial cables. 
This emulation mode was used only during the development of the system and is not intended for actual BOR measurements on the beamline.

Control and register access to all application firmware components are managed via an AXI-Lite interface \cite{axil}. A protocol conversion from the Processing System's (PS's) native AXI4 memory interface to an AXI-Lite interface is utilized.

%% file: 03_Software.tex
\section{Software}
\label{sec:software}
\begin{figure}[tb]
\centering
\includegraphics[width=1.0\textwidth]{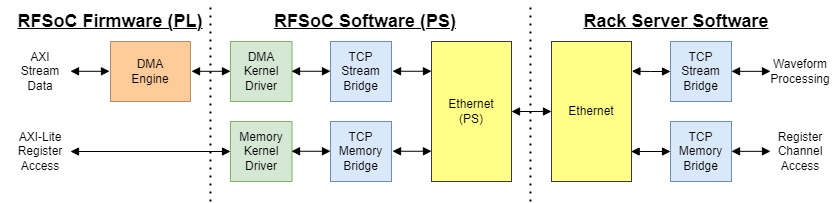}
\caption{\label{fig:sw_block_diagram} Block diagram of the software register access and waveform streaming.}
\end{figure}

The software register access and waveform streaming processes are illustrated in the block diagram in Figure \ref{fig:sw_block_diagram}. This communication involves three main components: RFSoC firmware, RFSoC software, and rack server software. Within the RFSoC firmware, waveforms are transmitted to the DMA engine and DMA kernel driver via the AXI Stream interface, while AXI-Lite register access is directed to the memory kernel driver. The memory kernel driver restricts user access to registers outside the application firmware context. The TCP stream bridge converts the received waveforms into TCP messages for Ethernet communication. Similarly, the TCP memory bridge translates the register access request and response messages between the memory kernel driver and the Ethernet TCP interface. On the rack server side, complementary TCP stream and TCP memory bridges translate the TCP messages back into the native run control software interfaces. Petalinux (6.1.30-xilinx-v2023.2) is used on the RFSoC for the Linux kernel.

The run control software used in this testing, known as "Rogue," was designed for both, rapid prototyping and experiment deployment. It can operate in either a Python/C++ hybrid mode or a C++-only mode, and is supported on x86-64, ARM32, and ARM64 architectures \cite{slaclab_rogue}. In this application, the Rogue software runs on both, the RFSoC and the rack server to utilize a common library for the TCP stream/memory bridge. The RFSoC's application register mapping within Rogue is defined using Python classes. The waveform receiver is also defined within a Python class, which converts the native byte array into a 16-bit numpy array for waveform data processing and analysis. Additionally, a GUI was developed on top of the Rogue software to assist in the live display of BPM waveforms and register diagnostics. 
The GUI software is based on the Python Display Manager (PyDM), a PyQt-based framework for building user interfaces for control systems \cite{slaclab_pydm}.


%% file: 04_Testing.tex
\section{Testing}
\label{sec:Testing}
\subsection{Overview}
We performed two types of tests on the RFSoC-based BOR using the two symmetrically arranged, vertical button BPM electrodes attached to the FB-BPM chamber of the LER. 
The first test was to evaluate the accuracy of the position measurement using a local beam bump, and the second test was to check whether the RFSoC-based BOR was working properly as a bunch-by-bunch detector using SuperKEKB's feedback system.

\subsection{Test using a local beam bump}
\label{sec:Test_local_beam_bump}
To evaluate the position resolution, we performed a test using a local beam bump.
We created a local, closed beam bump in the vertical plane using the steering magnets and measured the position of the vertically displaced beam using the RFSoC-based BOR.
This test was carried out on February 23, 2024 at the LER. 
With 97 bunches, the bunch spacing, approximately \SI{100}{\nano\second}, was wide enough so that the ringing or tail of a bunch signal did not interfere with other bunch signals.
The bunch intensity was approximately 5~nC resulting in an amplitude of approximately \SI{250}{\milli\volt} for the $\Sigma$ signal.
The beam position in the bump was monitored by nearby narrow-band BPM detectors, with a resolution of about \SI{2}{\micro\meter} \cite{SuperKEKBTDR}.
The narrow-band BPM detectors are used to measure the beam's closed orbit at a frequency of 0.25 Hz.
The bump position was aligned using two narrow-band BPMs placed on either side of the FB-BPM chamber. 
The narrow-band BPM located closer to the FB-BPM chamber was approximately \SI{1.6}{\meter} away.

\begin{figure}[hbt]
\centering
\includegraphics[width=0.6\textwidth]{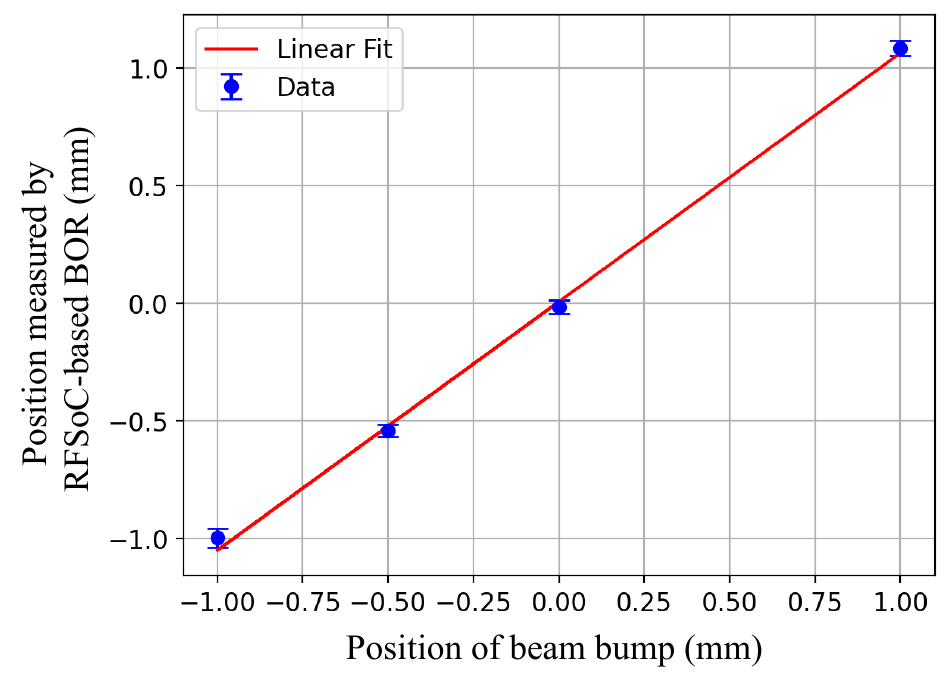}
\caption{\label{fig:resolution} Measurement results of the test using a local beam bump. The horizontal axis is the position of the beam bump, and the vertical axis is the vertical beam position measured by the RFSoC-based BOR. The blue plots is the measured point and the red line is the linear fit of these plots.}
\end{figure}

The measurement results of the test are summarized in Figure~\ref{fig:resolution}.
The vertical beam positions measured by the RFSoC-based BOR and calculated using equation~\ref{position_eq} (on the vertical axis in Figure~\ref{fig:resolution}) are compared with those measured by the narrow-band BPM (horizontal axis in Figure~\ref{fig:resolution}).
The good agreement is further demonstrated by the linear fit, shown as a red trace in Figure~\ref{fig:resolution}, confirming the good linearity of the RFSoC-based BOR. 
The fitting result is $y=(1.057\pm 0.029)x+(0.007\pm 0.020)$.
The standard deviation of each measurement point is approximately \SI{30}{\micro\meter}.
Since this value is much greater than the resolution of the narrow-band detector (\textasciitilde \SI{2}{\micro\meter}) used to set the bump position, we assume that the effect of the bump setting accuracy on the measurement results is negligible.
The achieved resolution of \SI{30}{\micro\meter} is comparable to that of the VME-BOR, meeting the target accuracy. 

At NSLS-II, the development of a bunch-by-bunch BPM using RFSoC has been underway, achieving a positional resolution of approximately \SI{15}{\micro\meter} \cite{ipac_nsls}.
The resolution of the RFSoC-BOR is currently worse than that.
This difference can be understood by the inability to fully utilize the ADC input range of the RFSoC.
Since we are using the standard mezzanine card and amplifier evaluation boards, an appropriate gain design has not yet been implemented.
Currently, we are developing a custom mezzanine card with surface-mount amplifiers offering higher gain, and we expect the accuracy to improve in the future.

\subsection{Test using the feedback system}
We validated the performance of the RFSoC-based BOR by comparing its position measurements with those from an established bunch-by-bunch detector. 
Specifically, we used the FB-BPM detector \cite{feedback} located in the Fuji straight section. 
The feedback system is based on the iGp12 made by Dimtel, inc. \cite{iGp12}, which includes digitizers, an FPGA that contains tap filters and other hardware, and is operated successfully at SuperKEKB, demonstrating a well-calibrated FB-BPM detector.
The bunch positions recorded by this detector have a resolution of about \SI{2}{\micro\meter}~\cite{SuperKEKBTDR}. 
By comparing them with the bunch positions measured by the RFSoC-based BOR, we tried to confirm that the RFSoC-based BOR was functioning properly as a bunch-by-bunch detector.
The FB-BPM detector achieves higher accuracy than the RFSoC-based BOR because it employs an analog circuit to subtract the offset caused by the closed orbit signals and amplifiers that fully utilize the ADC range.

For a demonstration of the bunch position measurements, in this test, we intentionally induced a bunch-by-bunch beam instability and monitored the bunch positions with both the RFSoC-based BOR and the FB-BPM detector simultaneously.
To induce a beam instability, we inverted the phase of the feedback kicker by 180 degrees while the beam was stored in the ring, which caused an intentional increase in the amplitude of the bunch position oscillations.
Since the FB-BPM detector and the RFSoC-based BOR used the button electrodes attached to the same chamber shown in Figure~\ref{fig:chamber}, both systems would in principle measure the same bunch positions.
This test was carried out on June 6, 2024 using the LER beam. 
The number of bunches was 393, the bunch intensity was approximately 5~nC, and the amplitude of the $\Sigma$ waveform was about \SI{250}{\milli\volt}. 
Since the bunch intensity was almost identical to that of the test using the local beam bump in Section \ref{sec:Test_local_beam_bump}, the resolution of the RFSoC-based BOR in this test should be approximately \SI{30}{\micro\meter}. 
The beam instability was induced in the vertical plane, with the FB-BPM detector and the RFSoC-based BOR both monitoring the vertical bunch displacements.

\begin{figure}[hbt]
\centering
\includegraphics[width=1.0\textwidth]{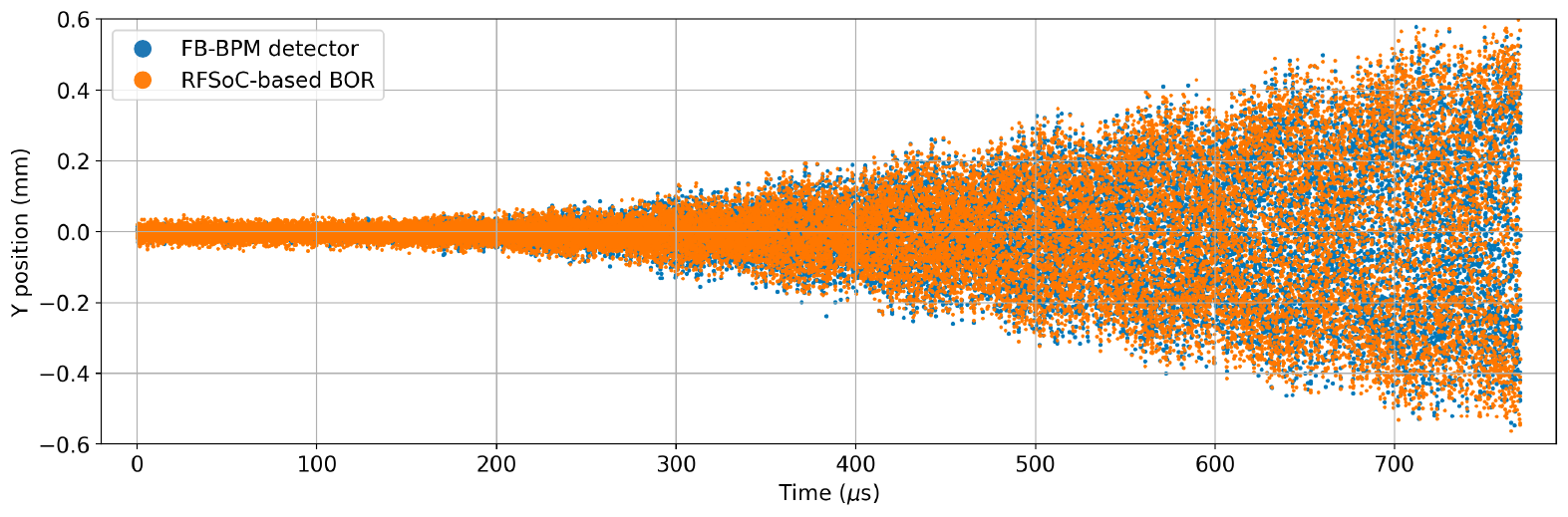}
\caption{\label{fig:igp_rfsoc} Measurement results of the test using feedback system. The blue plots represent the positions measured by the FB-BPM detector and the orange plots represent the positions measured by the RFSoC-based BOR. Each point represents one bunch. Bunch oscillations are increased by inverting the phase of the feedback kicker.}
\end{figure}

\begin{figure}[hbt]
\centering
\includegraphics[width=0.65\textwidth]{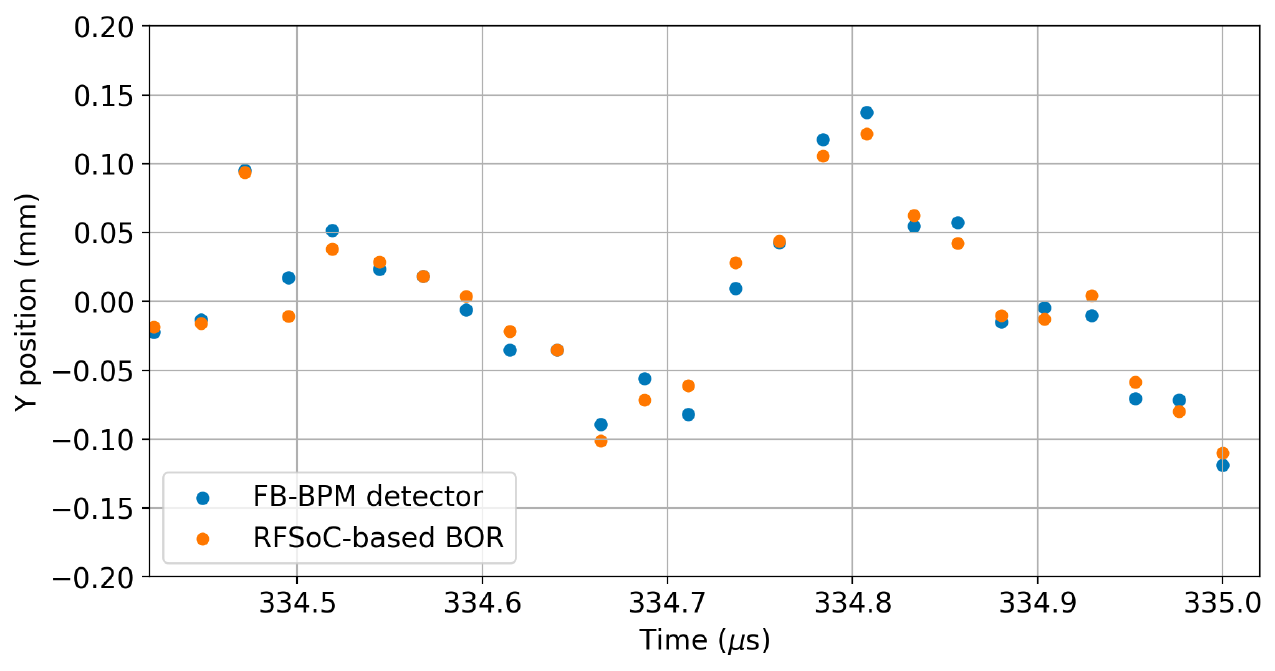}
\caption{\label{fig:igp_rfsoc_expand} Enlarged view of a specific part of Figure~\ref{fig:igp_rfsoc}}
\end{figure}

The measurement results are presented in Figure~\ref{fig:igp_rfsoc}. 
Each point represents the vertical position of one bunch vs. time.
The blue points represent the positions measured by the FB-BPM detector and the orange points represent the positions measured by the RFSoC-based BOR. 
A timing signal to invert the phase of the feedback kicker was received simultaneously by the FB-BPM detector and RFSoC-based BOR, and was used as the trigger to initiate the recording of the bunch-by-bunch measurements, such that the bunch position measurements by the FB-BPM detector and RFSoC-based BOR, shown in Figure~\ref{fig:igp_rfsoc}, were synchronized.
Starting at approximately \SI{150}{\micro\second} in Figure~\ref{fig:igp_rfsoc}, note how the amplitude of the bunch position starts to oscillate due to the phase inversion of the feedback kicker.
Comparing the FB-BPM detector and RFSoC-based BOR measurements, we see that the amplitude of the oscillations and the time it takes to increase are almost the same.
Additionally in Figure~\ref{fig:igp_rfsoc_expand}, an enlarged view of Figure~\ref{fig:igp_rfsoc}, we see that the measurement results at the single-bunch level agree well between the two types of monitors.
According to these results, we confirmed that the RFSoC-based BOR worked properly as a bunch-by-bunch detector.

%% file: 05_Analysis.tex
\section{Observation of Sudden Beam Loss}
\label{sec:Observation and analysis}

The RFSoC-based BOR was left connected to the FB-BPM chamber in the Fuji straight section after completing beam tests described in Section \ref{sec:Testing}. 
The beam operation of SuperKEKB resumed at the end of February 2024, and SBL observations using the RFSoC-based BOR started from thereon.

\begin{figure}[hbt]
\centering
\includegraphics[width=0.9\textwidth]{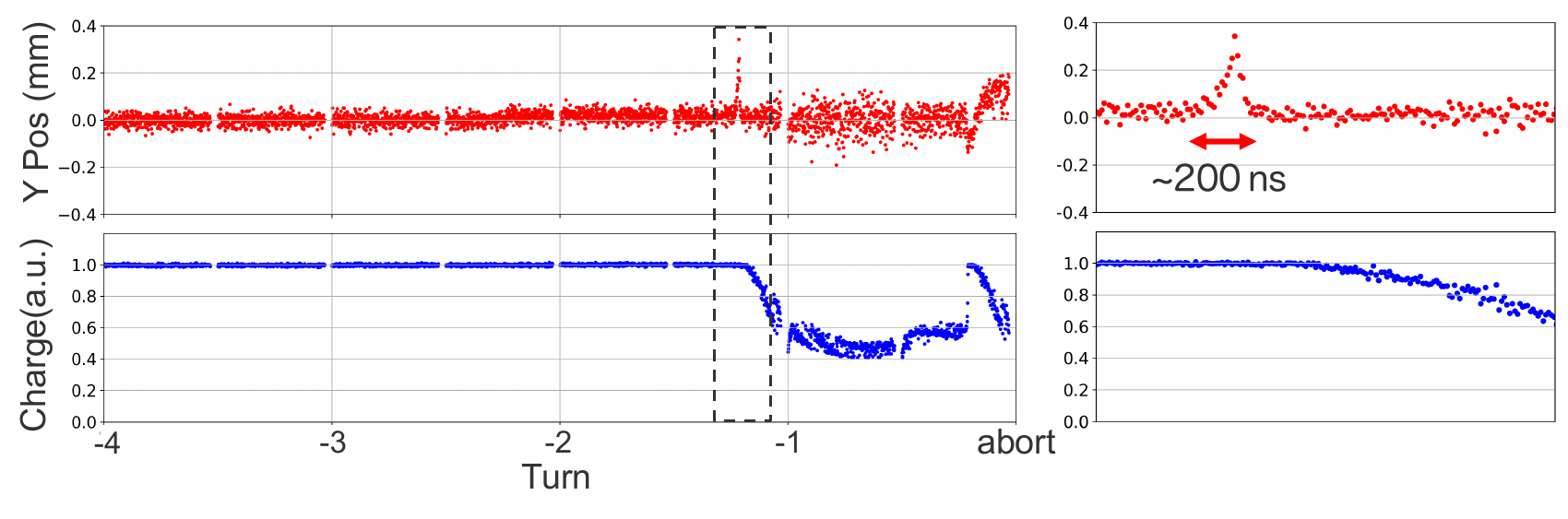}
\caption{\label{fig:first_SBL} The first SBL event recorded with RFSoC-based BOR. This SBL occurred at the LER on March 8, 2024. The number of bunches was 783, and the bunch intensity was 3.3~nC.  Left: Vertical position (upper) and charge (bottom) for the last 4 turns before the beam abort. One point corresponds to one bunch. Right: Enlarged view of the dotted line area in the figure on the left.}
\end{figure}

Figure~\ref{fig:first_SBL} shows the first SBL event recorded by the RFSoC-based BOR. 
This was the SBL that occurred at the LER on March 8, 2024. 
The number of bunches was 783 and the bunch intensity was 3.3~nC. 
The red dots in the upper half of this figure show the vertical bunch positions, and the blue dots in the lower half show the bunch charges. 
Each point represents one bunch. 
The first three turns of the recorded data were used to calculate the average position for each bunch, which was then subtracted from all data as the offset for each bunch. 
Also, the first three turns of the recorded data were used to calculate the average charge for each bunch, and then all the data was divided by this average charge to normalize all bunch charges to 1 for a situation of stable beam.
Figure~\ref{fig:first_SBL} (left) shows the bunch positions and bunch charges recorded by the RFSoC-based BOR for four turns before the beam abort, with the horizontal axis representing the number of turns before the beam abort.
There are two abort gaps in a single turn, each of approximately \SI{200}{\nano\second} duration, clearly visible in the plot as no data entry.

Figure~\ref{fig:first_SBL} (left) shows some bunches are kicked upwards approximately one turn before the beam is aborted, highlighted by the dotted-line area.
Figure~\ref{fig:first_SBL} (right) shows an enlarged view of that area, indicating a vertical change of the beam potions of \SI{0.4}{\milli\meter} for some bunches in the turn.
Immediately following, bunches started to rapidly lose intensity, up to half of the bunch charge was lost.
The figure on the right shows that only bunches within a timeframe of \SI{200}{\nano\second} duration were kicked significantly.
The charge of the kicked bunches did not decrease, but the charge of the following bunches started to decrease. 
The observation of an SBL event indicated that a sequence of bunches was kicked for an unknown reason and started to oscillate. Nevertheless, the cause of this kick could not be determined from these measurements alone.

\begin{figure}[hbt]
\centering
\includegraphics[width=1.0\textwidth]{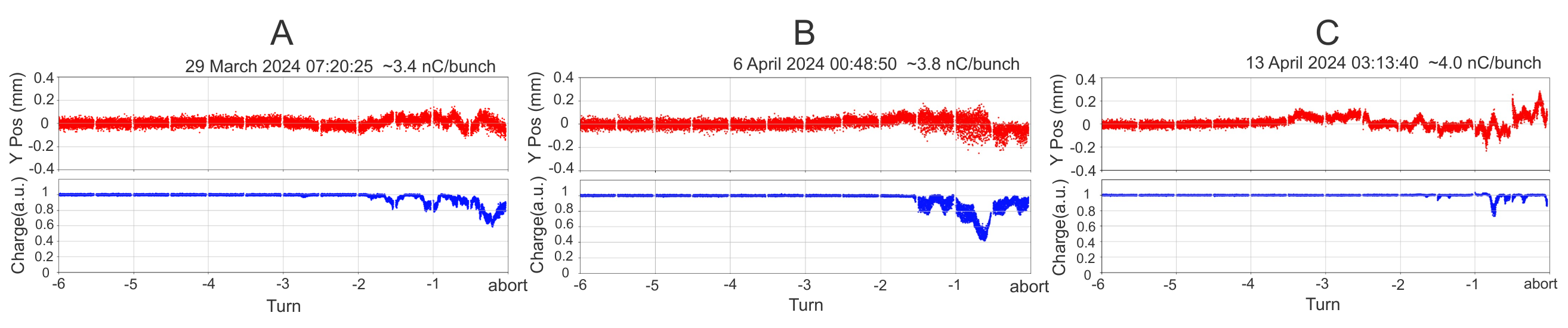}
\caption{\label{fig:some_SBL} Three examples of the SBLs recorded by RFSoC-based BOR. The upper plot shows the vertical position and the lower plot shows the bunch charge for the last 6 turns before the beam abort. From left to right,\\ 'A' represents an SBL that occurred on March 29, 2024, with 2346 bunches at a bunch intensity of 3.4~nC.\\ 'B' represents an SBL that occurred on April 6, 2024, with 2346 bunches at a bunch intensity of 3.8~nC.\\ 'C' represents an SBL that occurred on April 13, 2024, with 2346 bunches at a bunch intensity of 4.0~nC.}
\end{figure}

We continued to record more SBLs using the RFSoC-based BOR. 
Figure~\ref{fig:some_SBL} shows three examples of the recorded SBLs. 
Again, the upper trace shows the vertical position of the bunches, and the lower trace the normalized charges of the bunches, here for the last six turns before the beam abort.
These plots can be interpreted in the same way as in Figure~\ref{fig:first_SBL}.
From the left, the panel 'A' represents an SBL event on March 29, 2024, with 2346 bunches and a bunch intensity of 3.4~nC.
In this event, the bunch position displacement started approximately three turns before the beam abort, followed by the increase in the bunch oscillation amplitude.
The charge loss started approximately two turns before the beam abort. 
The panel 'B' represents an SBL event on April 6, 2024, with 2346 bunches at a bunch intensity of 3.8~nC.
In this event, the bunch position displacement started approximately three turns before the beam abort, followed by the increase in the bunch oscillation amplitude.
The bunch oscillation was different from those in 'A', and the bunch train seemed to be greatly disturbed.
The most significant loss in bunch intensity started approximately 1.5 turns before the beam abort.
The panel 'C' represents an SBL event on April 13, 2024, with 2346 bunches and a bunch intensity of 4~nC. 
In this event, the bunch position displacement started approximately four turns before the beam abort, and the oscillations started earlier compared to the other examples.
However, the amount of charge loss was not as great as in the other examples.

These measurement results show variations in the bunch position oscillations and charge losses during an SBL event.
Two types of SBLs were observed: those with rapidly developing oscillations, as shown in Figure~\ref{fig:first_SBL}, and those with more slowly developing oscillations, as shown in Figure~\ref{fig:some_SBL}.
These bunch oscillations were initiated when the bunch received some kind of kick, but these results indicate the characteristics of the kick, such as strength and location, vary for each SBL event, and it is still unclear what has caused these oscillations.
Therefore, it will be difficult to determine the cause of the SBL by analyzing the data just from a single BOR at a specific location in the ring.
We believe it is important to set up more BORs and install them in the SuperKEKB ring to better identify the root cause of the SBL.
This will enable us to cover a wider range of betatron phase space, possibly determining at what betatron phase the bunch oscillations begin, and how the oscillations develop. 
Furthermore, by using multiple BORs around the area where such a kick is likely to occur, it may be possible to directly pinpoint the origin of the SBL. 
When considering placing many BORs in the ring, the RFSoC-based BOR, which does not require a crate and is easy to carry, is expected to be a very powerful monitor. 
We plan to set up additional RFSoC-based BORs and place them around the entire ring to get a better insight into identifying the root cause of the SBL events.

%% file: 06_summary.tex
\section{Summary}
\label{sec:Summary}

The SuperKEKB accelerator aims to significantly increase the luminosity, but a phenomenon known as SBL has emerged as a challenge. This is a phenomenon in which an ampere-class stored beam is suddenly lost within a few turns (\textasciitilde a few tens of \si{\micro\second}). The SBL events can damage accelerator components and the Belle II detectors, and quench the superconducting focusing system, which hinders improvements in the integrated luminosity.

To address and analyze the SBL in detail, we developed a new BOR based on RFSoC technology. This BOR captures the bunch-by-bunch beam position just before the beam abort occurs. Our RFSoC-based BOR was verified to function effectively as a bunch-by-bunch monitor, achieving a position resolution of \SI{30}{\micro\meter}.
We have initiated SBL observations using the RFSoC-based BOR, successfully recording multiple SBL events. 
Though the RFSoC-based BOR in this study was installed near the FB-BPM detector for comparison purposes, in the future the RFSoC-based BOR will be relocated to another position along the ring and used for observing SBL.

The data recorded by the RFSoC-based BOR revealed a variety of positional oscillations and charge loss patterns during SBL events. Consequently, we believe that identifying the cause of SBL will require more BORs. In the future, we aim to deploy additional RFSoC-based BORs to cover a wider range of betatron oscillation phases, enabling a more comprehensive investigation of the SBL events. Furthermore, we intend to use BORs to identify the source of beam instability by strategically positioning them at locations suspected to be the origin of the SBL. We anticipate that the newly developed portable, high-speed bunch-by-bunch monitor will bring us closer to resolving the SBL problem.

\section{Acknowledgements}

R. Nomaru and G. Mitsuka work was supported by Japan Society for the Promotion of Science (JSPS) International Leading Research Grant Number JP22K21347. In addition, R. Nomaru work was supported by JSPS Core-to-Core Program (Grant Number: JPJSCCA20230004). L. Ruckman and R. Herbst work was supported by the U.S. Department of Energy, under contract number DE-AC02-76SF00515.
